\documentclass[lettersize,journal]{IEEEtran}
\usepackage{amsmath,amsfonts}
\usepackage{algorithmic}
\usepackage{array}
\usepackage[caption=false,font=normalsize,labelfont=sf,textfont=sf]{subfig}
\usepackage{textcomp}
\usepackage{stfloats}
\usepackage{latexsym}
\usepackage{amsmath} 
\usepackage{color}
\usepackage{amssymb} 
\usepackage{algorithm}
\usepackage{cleveref} 
\usepackage{algorithmic}
\usepackage{stfloats}
\usepackage{bm}
\usepackage{amsmath}

\usepackage{amssymb}
\usepackage{gensymb}
\usepackage{url}
\usepackage{amsmath}
\usepackage{pifont}
\usepackage{cite}
\usepackage{epsfig}
\usepackage{epstopdf}
\usepackage{graphicx}
\def\BibTeX{{\rm B\kern-.05em{\sc i\kern-.025em b}\kern-.08em
    T\kern-.1667em\lower.7ex\hbox{E}\kern-.125emX}}
\usepackage{balance}
\begin{document}
	\title{Intelligent Reflecting Surface-Aided Radar Spoofing}
	\author{Haozhe Wang,  Beixiong Zheng, \IEEEmembership{Senior Member,~IEEE,} Xiaodan Shao, \IEEEmembership{Member,~IEEE,}\\ and Rui Zhang, \IEEEmembership{Fellow,~IEEE}
	\thanks{H. Wang is with Shenzhen Research Institute of Big Data, School of Science and Engineering, The Chinese University of Hong Kong, Shenzhen, Guangdong, 518172, China (e-mail: haozhewang1@link.cuhk.edu.cn).}
	\thanks{B. Zheng is with School of Microelectronics, South China University of Technology, Guangzhou 511442, China, and also with Shenzhen Research Institute of Big Data, Shenzhen, 518172, China (e-mail: bxzheng@scut.edu.cn).}
	\thanks{X. Shao is with the Institute for Digital Communications, Friedrich-Alexander-University Erlangen-Nuremberg, 91054 Erlangen, Germany (e-mail: xiaodan.shao@fau.de).}
	\thanks{R. Zhang is with School of Science and Engineering, Shenzhen Research
	Institute of Big Data, The Chinese University of Hong Kong, Shenzhen,
	Guangdong 518172, China. He is also with the Department of Electrical and
	Computer Engineering, National University of Singapore, Singapore 117583
	(e-mail: elezhang@nus.edu.sg).}}
	\maketitle
\begin{abstract}
Electronic countermeasure (ECM) technology plays a critical role in modern electronic warfare, which can interfere with enemy radar detection systems by noise or deceptive signals. However, the conventional active jamming strategy incurs additional hardware and power costs and has the potential threat of exposing the target itself. To tackle the above challenges, we propose a new intelligent reflecting surface (IRS)-aided radar spoofing strategy in this letter, where IRS is deployed on the surface of a target to help eliminate the signals reflected towards the hostile radar to shield the target, while simultaneously redirecting its reflected signal towards a surrounding clutter to generate deceptive angle-of-arrival (AoA) sensing information for the radar. We optimize the IRS's reflection to maximize the received signal power at the radar from the direction of the selected clutter subject to the constraint that its received power from the direction of the target is lower than a given detection threshold. We first solve this non-convex optimization problem using the semidefinite relaxation (SDR) method and further propose a lower-complexity solution for real-time implementation. Simulation results validate the efficacy of our proposed IRS-aided spoofing system as compared to various benchmark schemes.
\end{abstract}
	
\begin{IEEEkeywords}
	Radar spoofing, intelligent reflecting surface (IRS), reflection optimization, angle-of-arrival (AoA) sensing.
\end{IEEEkeywords}

\section{Introduction}
\IEEEPARstart{E}{lectronic} countermeasure (ECM) technology has attracted growing attention with the development of electronic warfare due to its ability to evade/spoof enemy radar systems with noise or deceptive signals \cite{ECM_overview}. Conventional ECM can be categorized into two types: passive ECM and active ECM. Passive ECM involves the use of devices that do not emit electromagnetic (EM) waves to evade enemy radar detection, such as chaff clouds for target shielding and EM absorbing materials for target stealth. While active ECM disrupts enemy radar detection by actively transmitting EM waves, such as jamming \cite{tang2023techniques}. With the application of digital radio-frequency memory (DRFM) units, active ECM has become a predominant approach due to its ability to replicate and modify the received radar signals for retransmission, which is especially useful in deceiving the enemy radar of the target's distance and velocity \cite{kwak2009application}. 

However, passive ECM methods have limited flexibility and reconfigurability, while active ECM methods not only require costly hardware and additional power consumption, but also pose the potential risk by exposing the genuine direction of the target due to the broadcast nature of EM waves. Thus, it is imperative to develop high-performance radar spoofing strategies in a cost-effective manner.

On the other hand, intelligent reflecting surface (IRS) has emerged as a promising technology for enhancing wireless communication and sensing thanks to its low hardware cost and energy consumption \cite{zheng2022survey,qingqing}. By adjusting the reflection coefficients of low-cost passive reflecting elements, IRS is able to dynamically alter the reflected signal propagation to improve the communication/sensing performance. In particular, the use of IRS for achieving secure radar sensing has been recently studied in \cite{shaosss, 10474137, zheng2024intelligent}, where IRS is mounted on the surface of target to evade radar detection. In this context, IRS adjusts its reflection coefficients to suppress the echo signal from the target to one or more radars to greatly decrease their detection probability for the target. However, the IRS-aided radar spoofing still remains largely unexplored. In contrast to traditional ECM, IRS can provide flexible and real-time control over its reflected EM waves, which enables a new and cost-effective approach for radar spoofing.

Motivated by the above, we study a new IRS-aided radar spoofing strategy in this letter, where the target's reflected radar signal is diverted towards a selected surrounding clutter to create an echo signal with false angle-of-arrival (AoA) information for deceiving the adversarial radar, while simultaneously concealing the true target's AoA by suppressing the echo signal directly from the target to the radar. To this end, we formulate an optimization problem for designing the IRS's unit-modulus reflection coefficients to maximize the received echo signal power at the radar from the direction of the clutter subject to the constraint that its received power from the direction of the target is lower than a given detection threshold. To solve this non-convex optimization problem, we first apply the semidefinite relaxation (SDR) technique to obtain a high-quality suboptimal solution. Moreover, for the ease of real-time implementation, we propose a lower-complexity algorithm to solve the problem via the inequality transformation. Simulation results are provided to demonstrate the advantages of the proposed IRS-aided radar spoofing strategy as compared to various benchmark schemes.

\section{System Model}  
\begin{figure}[!t]
	\centering
	\includegraphics[width=0.34\textwidth]{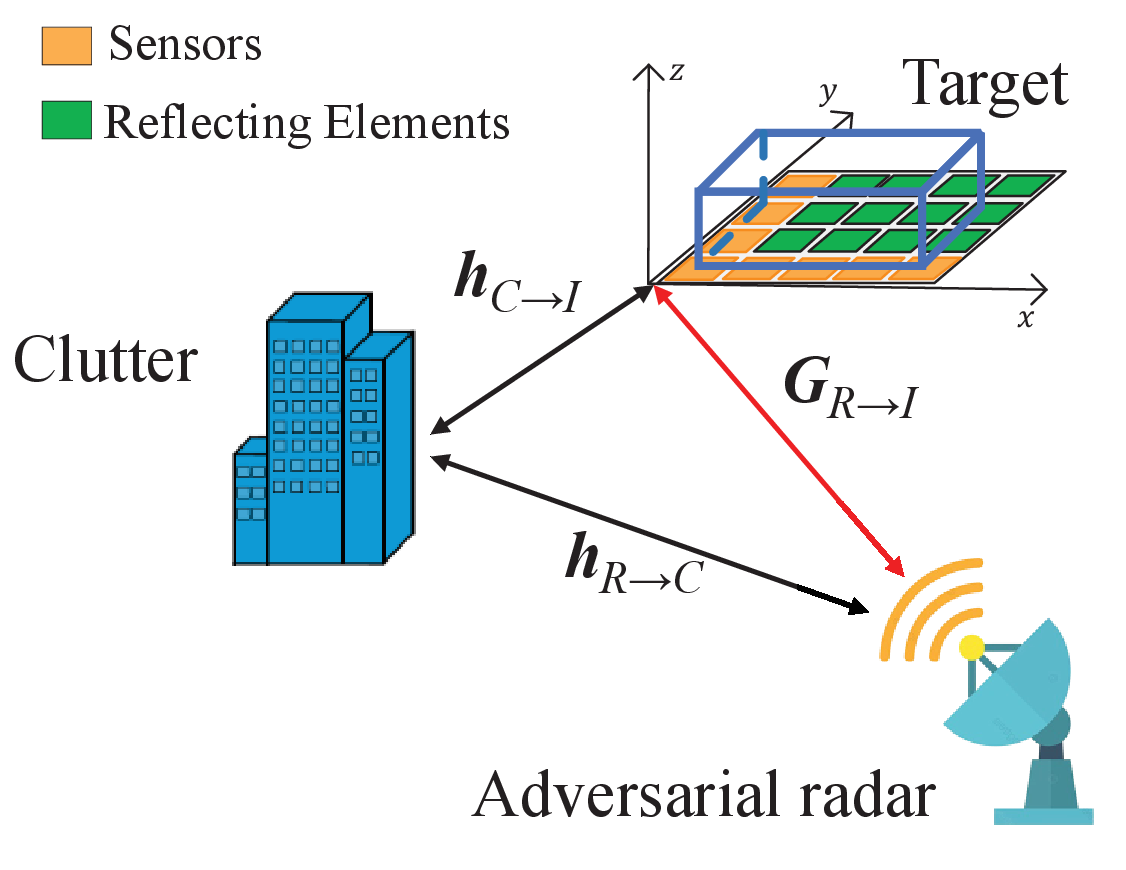}
	\caption{An IRS-aided radar spoofing system.}
	\label{fig:threshold111}
\end{figure}

As illustrated in Fig. \ref{fig:threshold111}, we consider an IRS-aided radar spoofing system mounted on a moving target (e.g., aircraft), where an IRS is installed on the target's surface to spoof the adversarial radar detection with deceptive angle sensing information. Without loss of generality, we consider that the IRS is a uniform planar array (UPA), which is composed of $N \triangleq N_x\times N_y$ reflecting elements, where $N_x$ and $N_y$ are the number of reflecting elements along  $x-$ and $y-$axes, respectively. The reflection coefficients of the IRS for time epoch $t$ are denoted by the matrix $\boldsymbol{\Theta}^{[t]}=\operatorname{diag}\left(\beta^{[t]}_1 e^{j \theta_1^{[t]}}, \ldots, \beta_n^{[t]} e^{j \theta_N^{[t]}}\right) \in \mathbb{C}^{N \times N}$, where $\beta_n^{[t]}\in\left\{0,1\right\}$ and $\theta_n^{[t]} \in [0,2\pi)$ with $n=1,\ldots,N$ are the ON/OFF amplitude and phase shift of the $n$-th element for time epoch $t$, respectively. To facilitate the estimation of AoAs and powers of radar probing signals from different directions at the IRS/target, a sensing array consisting of $L=L_x+L_y-1$ sensing devices is equipped at the edges of IRS (see Fig. \ref{fig:threshold111}). Specifically, this semi-passive IRS operates in two alternating modes, i.e., sensing mode and reflection mode. In the sensing mode, all reflecting elements are switched OFF (i.e., $\beta_n^{[t]}=0, \forall n\in\{1,\ldots,N\}$), indicating that the incident radar signals are fully absorbed by the reflecting elements (similar to EM stealth materials). Meanwhile, the sensing array estimates the AoAs of the incident radar signals and their powers.\footnote{The method of estimating the signal AoAs and powers from the radar and clutter directions at the IRS sensing array can be referred to \cite{zheng2023intelligent}.} In the subsequent reflection mode, all reflecting elements are switched ON (i.e., $\beta_n^{[t]}=1,\forall n\in\{1,\ldots,N\}$) to adjust phase shifts for radar spoofing based on the parameters estimated in the sensing mode. Moreover, we assume a mono-static full-duplex adversarial radar equipped with a UPA consisting of $M \triangleq M_x \times M_y$ transmit/receive antennas. 
\subsection{Channel Model} 
Let $\boldsymbol G^{[t]}_{R\rightarrow I}\in\mathbb{C}^{N\times M}$, $\boldsymbol h_{R\rightarrow C}^{[t]}\in\mathbb{C}^{M\times 1}$, and $\boldsymbol h_{C\rightarrow I}^{[t]} \in \mathbb{C}^{M\times 1}$ denote the equivalent baseband channels for the radar$\rightarrow$IRS/target, radar$\rightarrow$clutter, and clutter$\rightarrow$IRS/target for time epoch $t$, respectively. Due to the high altitude of the aerial target, the propagation channels among the radar, IRS/target, and clutter can be characterized by the far-field line-of-sight (LoS) model. Let $\boldsymbol{e}(\phi,N_l)$ denote the one-dimensional (1D) steering vector function for a generic uniform linear array (ULA), which is defined as
\begin{equation}
	\boldsymbol{e}(\phi,N_l) \triangleq \left[1, e^{-j \frac{2\pi d}{\lambda} \phi}, \cdots, e^{-j \frac{2\pi d}{\lambda}(N_l-1) \phi}\right]^T,
\end{equation}
where $j=\sqrt{-1}$ denotes the imaginary unit, $\phi$ denotes the constant phase-shift difference between the signals at two adjacent antennas/elements, $d$ denotes the distance between two adjacent antennas/elements, $\lambda$ denotes the signal wavelength, and $N_l$ denotes the number of antennas/elements of the ULA. The IRS and target share the same AoAs/angles-of-departure (AoDs) for the radar signals given the same reference point. We denote the elevation and azimuth AoA pair of the received radar signal at the IRS/target from the direction of the radar as ($\vartheta_{R \rightarrow I}^{[t]}$, $\eta_{R\rightarrow I}^{[t]}$) and that from the direction of the clutter as ($\vartheta_{C\rightarrow I}^{[t]}$, $\eta_{C \rightarrow I}^{[t]}$). Similarly, we denote the AoD pair from the adversarial radar towards the direction of the clutter as ($\vartheta_{R \rightarrow C}^{[t]}$, $\eta_{R \rightarrow C}^{[t]}$). Accordingly, we let $\boldsymbol{a}_I(\vartheta_{R\rightarrow I}^{[t]}$, $\eta_{R\rightarrow I}^{[t]})$ and $\boldsymbol{a}_I(\vartheta_{C \rightarrow I}^{[t]}$, $\eta_{C \rightarrow I}^{[t]})$ denote the two-dimensional (2D) steering vectors at the IRS for the incident signals from the direction of the radar and clutter, respectively. We also let $\boldsymbol{a}_R(\vartheta_{R \rightarrow I}^{[t]}$, $\eta_{R \rightarrow I}^{[t]})$ and $\boldsymbol{a}_R(\vartheta_{R \rightarrow C}^{[t]}$, $\eta_{R \rightarrow C}^{[t]})$ denote the 2D steering vectors at the radar for transmitting probing signals towards the direction of the IRS/target and clutter, respectively. Accordingly, the 2D steering vector can be expressed as the Kronecker product of two 1D steering vectors under the UPA model. For example, the 2D steering vector at the IRS/target from the direction of the radar can be expressed as
\begin{align}
	\boldsymbol{a}_I(\vartheta_{R \rightarrow I}^{[t]}, \eta_{R \rightarrow I}^{[t]}) &= \boldsymbol{e}\left( \cos \left(\eta_{R \rightarrow I}^{[t]}\right) \sin \left(\vartheta_{R \rightarrow I}^{[t]}\right), N_x\right) \nonumber\\
	& \quad \otimes \boldsymbol{e}\left( \sin \left(\eta_{R \rightarrow I}^{[t]}\right) \sin \left(\vartheta_{R \rightarrow I}^{[t]}\right), N_y\right). \label{eq:eq5}
\end{align}

The other 2D steering vectors $\boldsymbol{a}_I(\vartheta_{C \rightarrow I}^{[t]}$, $\eta_{C \rightarrow I}^{[t]})$,  $\boldsymbol{a}_R(\vartheta_{R \rightarrow I}^{[t]}$, $\eta_{R \rightarrow I}^{[t]})$, and $\boldsymbol{a}_R(\vartheta_{R \rightarrow C}^{[t]}$, $\eta_{R \rightarrow C}^{[t]})$ can be similarly defined as above, which are omitted for brevity.

Thus, the far-field LoS channel for the radar$\rightarrow$IRS/target $\boldsymbol G^{[t]}_{R\rightarrow I}\in\mathbb{C}^{N\times M}$ can be expressed as
\begin{align}  
	\boldsymbol {G}^{[t]}_{R\rightarrow I}  &= \rho^{[t]}_{R \rightarrow I} \boldsymbol{a}_I(\vartheta_{R \rightarrow I}^{[t]}, \eta_{R \rightarrow I}^{[t]}) \boldsymbol{a}^T_R(\vartheta_{R \rightarrow I}^{[t]}, \eta_{R \rightarrow I}^{[t]}), \label{eq:eq11}
\end{align}
where $\rho^{[t]}_{R \rightarrow I}=\frac{\sqrt{\alpha}}{d^{[t]}_{RI}}e^{j\frac{2\pi d^{[t]}_{RI}}{\lambda}}$ is the equivalent complex-valued path gain between them for time epoch $t$ with $\alpha$ being the reference path gain at a distance of 1 meter (m) and $d^{[t]}_{RI}$ denoting the distance between the radar and IRS/target for time epoch $t$. 

Similarly, the radar$\rightarrow$clutter channel $\boldsymbol h_{R\rightarrow C}^{[t]}\in\mathbb{C}^{M\times 1}$ and the clutter$\rightarrow$IRS channel $\boldsymbol h_{C\rightarrow I}^{[t]}\in\mathbb{C}^{M\times 1}$ can be respectively written as
\begin{align}  
\boldsymbol h_{R\rightarrow C}^{[t]} &= \rho^{[t]}_{R \rightarrow C} \boldsymbol{a}_R(\vartheta_{R \rightarrow C}^{[t]}, \eta_{R \rightarrow C}^{[t]}), \label{eq:eq13}\\
\boldsymbol h_{C\rightarrow I}^{[t]} &= \rho^{[t]}_{C \rightarrow I} \boldsymbol{a}_I(\vartheta_{C \rightarrow I}^{[t]}, \eta_{C \rightarrow I}^{[t]}), \label{eq:eq14}
\end{align}
where $\rho^{[t]}_{R \rightarrow C}=\frac{\sqrt{\kappa\alpha}}{d^{[t]}_{RC}}e^{j\frac{2\pi d^{[t]}_{RC}}{\lambda}}$ and $\rho^{[t]}_{C \rightarrow I}=\frac{\sqrt{\kappa\alpha}}{d^{[t]}_{CI}}e^{j\frac{2\pi d^{[t]}_{CI}}{\lambda}}$ are the corresponding equivalent complex-valued path gains with $\kappa$ denoting the radar cross section (RCS) of the clutter. In the above, $d^{[t]}_{RC}$ ($d^{[t]}_{CI}$) represents the distance between the radar (IRS) and clutter for time epoch $t$.

For the purpose of exposition, we assume that all the involved LoS channels have approximately constant amplitudes during each channel coherence interval and the channel reciprocity holds for each link in its forward and reverse directions. However, to characterize the effect of small variations in $d^{[t]}_{RI}$ and $d^{[t]}_{CI}$ due to local perturbation in the target's position on the signal phases $\nu^{[t]}_{R\rightarrow I}\triangleq \frac{2\pi d^{[t]}_{RI}}{\lambda} $ and $\nu^{[t]}_{C\rightarrow I}\triangleq \frac{2\pi d^{[t]}_{CI}}{\lambda}$, we model $\nu^{[t]}_{R\rightarrow I}$ and $\nu^{[t]}_{C\rightarrow I}$ as independent and uniformly distributed random variables in $[0,2\pi)$ over time epoch $t$\cite{liu2019comp}.
\subsection{Signal Model}
\begin{figure*}[ht] 
	\begin{equation}
		\setlength{\belowdisplayskip}{0pt}
		\label{eq:long}
		\boldsymbol{Y}^{[t]} =\underbrace{\left[\boldsymbol{G}^{[t]}_{R\rightarrow I}+\boldsymbol{h}^{[t]}_{C\rightarrow I}\left(\boldsymbol{h}^{[t]}_{R\rightarrow C}\right)^T\right]^T\boldsymbol{\Theta}\left[\boldsymbol{G}^{[t]}_{R\rightarrow I}+\boldsymbol{h}^{[t]}_{C\rightarrow I}\left(\boldsymbol{h}^{[t]}_{R\rightarrow C}\right)^T\right]\boldsymbol{S}^{[t]}}_{\text{Reflected by IRS}}+ \underbrace{\left[\boldsymbol{h}^{[t]}_{R\rightarrow C} \left(\boldsymbol{h}^{[t]}_{R\rightarrow C}\right)^T\right]\boldsymbol{S}^{[t]}}_{\text{Background}}+ \boldsymbol{Z}^{[t]},
	\end{equation}
	\begin{equation}
		\setlength{\belowdisplayskip}{0pt}
		\label{eq:longer}
		\boldsymbol{\tilde{Y}}^{[t]} = \underbrace{\left(\boldsymbol{G}^{[t]}_{R\rightarrow I}\right)^T \boldsymbol{\Theta}\left[ \boldsymbol{G}^{[t]}_{R\rightarrow I}+\boldsymbol{h}^{[t]}_{C\rightarrow I}\left(\boldsymbol{h}^{[t]}_{R\rightarrow C}\right)^T\right]\boldsymbol{S}^{[t]}}_{\text{ Received echo signal from the direction of the IRS/target}} +\underbrace{\boldsymbol{h}^{[t]}_{R\rightarrow C}\left(\boldsymbol{h}^{[t]}_{C\rightarrow I}\right)^T \boldsymbol{\Theta}\left[ \boldsymbol{G}^{[t]}_{R\rightarrow I} +\boldsymbol{h}^{[t]}_{C\rightarrow I}\left(\boldsymbol{h}^{[t]}_{R\rightarrow C}\right)^T\right]\boldsymbol{S}^{[t]}}_{\text{ Received echo signal from the direction of the clutter}}+ \boldsymbol{Z}^{[t]}.
	\end{equation}
	\hrulefill
\end{figure*}
Let $\boldsymbol{S}^{[t]} = \left[\boldsymbol{s}^{[t]}_1,\cdots,\boldsymbol{s}^{[t]}_K\right]$ denote the transmitted radar waveform for time epoch $t$ with $\mathbb{E}\left\{\boldsymbol{S}^{[t]}(\boldsymbol{S}^{[t]})^H\right\}=\textbf{I}_M$ and $K$ indicating the number of transmitted samples.
Based on the channel model, the received echo signals at the radar for time epoch $t$ can be expressed as in \eqref{eq:long} at the top of this page, where  $\boldsymbol{Z}^{[t]} \in \mathbb{C}^{M \times K} $ is an additive white Gaussian noise (AWGN) matrix with independent elements of zero mean and variance $\sigma^2$. In the following, we omit the second term in \eqref{eq:long} to further derive \eqref{eq:longer} since the background echo signal between the radar and clutter can be removed prior to target detection by the radar, regardless of the presence of any target.

In radar detection, the performance of AoA estimation mainly relies on the received echo signal power from the target's direction. A larger average received signal power from the target results in a higher signal-to-noise ratio (SNR) at the radar, thus leading to more accurate AoA estimation. Consequently, we adopt the average received signal power as the performance metric to evaluate the effectiveness of the IRS-aided radar spoofing system. Moreover, we drop the time epoch index $[t]$ for brevity in the following since we consider the average received power at the IRS/target or radar over multiple time epochs during each channel coherence interval.

Accordingly, the average received signal power at the adversarial radar from the direction of the IRS/target can be expressed as 
\begin{align}
	\setlength{\belowdisplayskip}{0pt}
	\overline{P}_{T} &= \mathbb{E}\left[\left\|\boldsymbol{G}^T_{R\rightarrow I} \boldsymbol{\Theta} \left(\boldsymbol{h}_{C\rightarrow I}\boldsymbol{h}^T_{R\rightarrow C}+\boldsymbol{G}_{R\rightarrow I}\right)\right\|^2_F\right] \nonumber\\
	&= \mathbb{E}\left[\left\|\boldsymbol{G}^T_{R\rightarrow I} \boldsymbol{\Theta} \boldsymbol{h}_{C\rightarrow I }\boldsymbol{h}_{R\rightarrow C}^T\right\|^2_F\right] \nonumber\\
	&\quad+\mathbb{E}\left[\left\|\boldsymbol{G}_{R\rightarrow I}^T \boldsymbol{\Theta} \boldsymbol{G}_{R\rightarrow I}\right\|^2_F\right], \label{eq:los}
\end{align}	
where the expectation is taken over time epochs. The above equation holds since the signal phase $\nu_{R\rightarrow I}$  in $\boldsymbol{G}_{R\rightarrow I}$ is independent of $\nu_{C\rightarrow I}$ in $\boldsymbol{h}_{C\rightarrow I}$. Therefore, the cross terms in \eqref{eq:los} are zero after taking the expectation due to $\mathbb{E}\left[e^{\pm j \nu_{R\rightarrow I}}\right]=0$ and $\mathbb{E}\left[e^{\pm j \nu_{C\rightarrow I}}\right]=0$. In particular, by substituting \eqref{eq:eq11}, \eqref{eq:eq13}, \eqref{eq:eq14} into \eqref{eq:los}, $\overline{P}_{T}$ can be further simplified as
 \begin{align}
 \overline{P}_{T}&= Q_RQ_C\left|\boldsymbol{g}^H\boldsymbol{\theta}\right|^2 + Q_R^2\left|\boldsymbol{v}^H\boldsymbol{\theta}\right|^2, 	\label{eq:lo1s}
\end{align}
where 
 \begin{align}
\boldsymbol{g}^H&\triangleq \boldsymbol{a}^T_I(\vartheta_{R \rightarrow I}, \eta_{R \rightarrow I}) \odot \boldsymbol{a}^T_I(\vartheta_{C\rightarrow I}, \eta_{C \rightarrow I}), \\
\boldsymbol{v}^H&\triangleq \boldsymbol{a}^T_I(\vartheta_{R \rightarrow I}, \eta_{R \rightarrow I}) \odot \boldsymbol{a}^T_I(\vartheta_{R\rightarrow I}, \eta_{R \rightarrow I}),
\end{align}
denote the cascaded steering vectors at the IRS, $\boldsymbol{\theta} \triangleq \left[\beta_1e^{j \theta_1}, \ldots, \beta_ne^{j \theta_N}\right]^T \in \mathbb{C}^{N \times 1}$ is the IRS reflection vector, and
\begin{align}
Q_R&=\left|\rho_{R \rightarrow I}\right|^2\left\|\boldsymbol{a}^T_R(\vartheta_{R \rightarrow I}, \eta_{R \rightarrow I})\right\|_F^2,\\ Q_C&=\left|\rho_{R \rightarrow C}\rho_{C \rightarrow I}\right|^2 \left\|\boldsymbol{a}^T_R(\vartheta_{R \rightarrow C}, \eta_{R \rightarrow C})\right\|_F^2,
\end{align}
are the received signal powers at the IRS sensor array from the direction of the radar and clutter, respectively.

After removing the background echo signal, the average received signal power at the adversarial radar from the direction of the clutter can be expressed as 
\begin{align}
	\overline{P}_{C} &= \mathbb{E}\left[\left\|\boldsymbol{h}_{R\rightarrow C}\boldsymbol{h}_{C\rightarrow I}^T \boldsymbol{\Theta} \left(\boldsymbol{h}_{C\rightarrow I}\boldsymbol{h}_{R\rightarrow C}^T + \boldsymbol{G}_{R\rightarrow I}\right)\right\|^2_F\right] \nonumber\\
	&= \mathbb{E}\left[\left\|\boldsymbol{h}_{R\rightarrow C}\boldsymbol{h}_{C\rightarrow I}^T \boldsymbol{\Theta}  \boldsymbol{h}_{C\rightarrow I}\boldsymbol{h}_{R\rightarrow C}^T\right\|^2_F\right] \nonumber\\
	&\quad + \mathbb{E}\left[\left\|\boldsymbol{h}_{R\rightarrow C}\boldsymbol{h}_{C\rightarrow I}^T \boldsymbol{\Theta} \boldsymbol{G}_{R\rightarrow I}\right\|^2_F\right].
\end{align}
Similar to \eqref{eq:lo1s}, $\overline{P}_{C}$  can be further simplified as 
\begin{align}
	\overline{P}_{C} &=Q_C^2\left|\boldsymbol{r}^H\boldsymbol{\theta}\right|^2 + Q_CQ_R\left|\boldsymbol{g}^H\boldsymbol{\theta}\right|^2,
\end{align}
where $\boldsymbol{r}^H \triangleq \boldsymbol{a}^T_I(\vartheta_{C \rightarrow I}, \eta_{C \rightarrow I}) \odot \boldsymbol{a}^T_I(\vartheta_{C\rightarrow I}, \eta_{C \rightarrow I})$ denotes the cascaded steering vector at the IRS.
\section{Problem Formulation and Solution}
We aim to maximize $\overline{P}_{C}$ for effectively diverting the echo signals from the IRS/target to clutter and then radar's receiver when the IRS is switched ON (i.e., $\beta_n^{[t]}=1, \forall n\in\{1,\ldots,N\}$). Meanwhile, it is required that the target remains stealthy under the constraint that $\overline{P}_{T}$ is lower than a given detection threshold, $\gamma$. Therefore, the corresponding optimization problem is formulated as
\begin{subequations}\label{prob_original}
	\begin{align}
		(\mathcal{P}_1):~ \mathop{\max}\limits_{\boldsymbol{\theta}} ~& Q_C^2\left|\boldsymbol{r}^H\boldsymbol{\theta}\right|^2 + Q_CQ_R\left|\boldsymbol{g}^H\boldsymbol{\theta}\right|^2
		\label{ooP1a}\\
		\mathrm{s.t.} ~ & Q_R^2\left|\boldsymbol{v}^H\boldsymbol{\theta}\right|^2 + Q_CQ_R\left|\boldsymbol{g}^H\boldsymbol{\theta}\right|^2 \leq \gamma,\label{ooP1b}\\
		~ & |\theta_n|= 1, \quad n=1,\cdots,N. \label{ooP1aaaa}
	\end{align}
\end{subequations}
\subsection{SDR-based Solution}
\label{sec:sdr}
Problem $(\mathcal{P}_1)$ is non-convex due to the unit-modulus constraint \eqref{ooP1aaaa}. We first consider applying the SDR method \cite{luo2010semidefinite} to solve this problem suboptimally. By relaxing the rank-one constraint on $\boldsymbol{\Theta}$, $(\mathcal{P}_1)$  can be reformulated as
\begin{subequations}\label{prob_original2}
	\begin{align}
		(\mathcal{P}_2):~ \mathop{\max}\limits_{\boldsymbol{\Theta}} ~& \operatorname{Tr}(\mathbf{A} \boldsymbol{\Theta})
		\label{ooP2a}\\
		\mathrm{s.t.} ~ & \operatorname{Tr}(\mathbf{B} \boldsymbol{\Theta}) \leq \gamma,\\
		~ & \boldsymbol{\Theta}_{n, n}=1, \forall n=1, \cdots, N, \\
		~ & \boldsymbol{\Theta} \succeq 0,
	\end{align}
\end{subequations}
where $\boldsymbol{\Theta}=\boldsymbol{\theta}\boldsymbol{\theta}^H$, $\mathbf{A}=Q_C^2\boldsymbol{r}\boldsymbol{r}^H+Q_CQ_R\boldsymbol{g}\boldsymbol{g}^H$, and $\mathbf{B}=Q_R^2\boldsymbol{v}\boldsymbol{v}^H+Q_CQ_R\boldsymbol{g}\boldsymbol{g}^H$. Note that $(\mathcal{P}_2)$  is a standard convex semidefinite programming (SDP), which can be solved by the CVX toolbox. After solving $(\mathcal{P}_2)$, the Gaussian randomization technique can be applied to find a high-quality suboptimal solution to $(\mathcal{P}_1)$.
\subsection{Low-Complexity Solution}
Next, we propose an alternative approach for solving problem $(\mathcal{P}_1)$ with a lower complexity than the SDR-based solution. This solution applies the majorization-minimization (MM) technique to obtain the closed-form solution for the phase shifts of IRS in an iterative manner. First, we rewrite $(\mathcal{P}_1)$ equivalently as
\begin{subequations}\label{prob_original3}
	\begin{align}
		(\mathcal{P}_3):~ \mathop{\max}\limits_{\boldsymbol{\theta}} ~& \boldsymbol{\theta}^H \mathbf{A} \boldsymbol{\theta}
		\label{ooP3a}\\
		\mathrm{s.t.} ~ &\boldsymbol{\theta}^H \mathbf{B} \boldsymbol{\theta} \leq \gamma,\label{P3_con}\\
		~ & |\theta_n|= 1, \quad n=1,\cdots,N, \label{ooP3aaaa}
	\end{align}
\end{subequations}
where $\mathbf{A}$ and $\mathbf{B}$ are defined in $(\mathcal{P}_2)$. By applying the first-order Taylor expansion, \eqref{ooP3a} is lower-bounded by
\begin{align}
	\boldsymbol{\theta}^H \mathbf{A} \boldsymbol{\theta} \geq 2\Re\left\{\boldsymbol{\theta}^H \mathbf{d}\right\} - c_1,
\end{align}
where $\mathbf{d}=\mathbf{A}\tilde{\boldsymbol{\theta}}$, $c_1=\tilde{\boldsymbol{\theta}}^H\mathbf{A}\tilde{\boldsymbol{\theta}}$, and $\tilde{\boldsymbol{\theta}}$ is the phase shifts vector obtained in the previous iteration. Next, we find an upper bound for the left-hand side of constraint \eqref{P3_con} based on the MM inequality transformation \cite{sun2016majorization}, which is given by
\begin{align}
	\boldsymbol{\theta}^H \mathbf{B} \boldsymbol{\theta} \leq \boldsymbol{\theta}^H \mathbf{M} \boldsymbol{\theta} + 2 \Re \left\{\boldsymbol{\theta}^H\left(\mathbf{B} - \mathbf{M}\right)\tilde{\boldsymbol{\theta}}\right\}+\tilde{\boldsymbol{\theta}}^H\left(\mathbf{M}-\mathbf{B}\right)\tilde{\boldsymbol{\theta}} \label{mm},
\end{align}
where $\mathbf{M}\succeq\mathbf{B}$. In this letter, we set $\mathbf{M}=\lambda_{max}\left(\mathbf{B}\right)\mathbf{I}_N$ since $\boldsymbol{\theta}^H\boldsymbol{\theta}=N$. We thus obtain $\boldsymbol{\theta}^H \mathbf{M} \boldsymbol{\theta}=N\lambda_{max}(\mathbf{B})$ with $\lambda_{max}(\mathbf{B})$ denoting the largest eigenvalue of matrix $\mathbf{B}$. By substituting \eqref{mm} into \eqref{P3_con}, \eqref{P3_con} can be rewritten as
\begin{align}
	 2 \Re \left\{\boldsymbol{\theta}^H\left(\mathbf{M}-\mathbf{B}\right)\tilde{\boldsymbol{\theta}}\right\}\geq c_2, \label{mm2}
\end{align}
where $c_2=N\lambda_{max}(\mathbf{B})+\tilde{\boldsymbol{\theta}}^H\left(\mathbf{M}-\mathbf{B}\right)\tilde{\boldsymbol{\theta}}-\gamma$. However, the problem is still non-convex due to the constraint \eqref{ooP3aaaa}. It is worth noting that a non-negative $\mu$ always exists, such as $(\mathcal{P}_3)$ can be reformulated into the following equivalent problem,
\begin{subequations}\label{prob_original4}
	\begin{align}
	(\mathcal{P}^{'}_3):~ \mathop{\max}\limits_{\boldsymbol{\theta}} ~& 2\Re\left\{\boldsymbol{\theta}^H \mathbf{d}\right\}+2\mu \Re \left\{\boldsymbol{\theta}^H\mathbf{e}\right\}\\
	\mathrm{s.t.} ~ & |\theta_n|= 1, \quad n=1,\cdots,N, \label{ooP333aaaa}
\end{align}
\end{subequations}
where $\mathbf{e}=\left(\mathbf{M}-\mathbf{B}\right)\tilde{\boldsymbol{\theta}}$. It is obvious that the objective function attains its maximum only when the phases of $\boldsymbol{\theta}$ and $\mathbf{d}+\mu\mathbf{e}$ are equal. Thus, the optimal solution of problem $(\mathcal{P}^{'}_3)$ is
\begin{align}
\boldsymbol{\theta}_{\text {opt }}(\mu)=\left[e^{j \arg \left(d_1+\mu e_1\right)}, e^{j \arg \left(d_2+\mu e_2\right)}, \cdots, e^{j \arg \left(d_N+\mu e_N\right)}\right]^T
\end{align}
where $\mu$ is an unknown dual variable. Substituting $\boldsymbol{\theta}_{\text {opt}}(\mu)$ into \eqref{mm2}, $\mu$ can be obtained according to the complementary slackness condition $\mu\left[2 \Re\left\{\boldsymbol{\theta}^H \mathbf{e}\right\}-c_2\right]=0$. To solve this equation, we consider the following two cases: 

(1) $\mu=0:$ In this case, $\boldsymbol{\theta}_{\text {opt }}(0)=e^{j \arg (\mathbf{d})}$ has to satisfy constraint \eqref{mm2}; otherwise, $\mu > 0$.

(2) $\mu>0:$ The complementary slackness condition holds if and only if $2 \Re\left\{\theta^H \mathbf{e}\right\}=c_2$ in this case. Thus, $\mu$ can be obtained via the bisection search since $2 \Re\left\{\theta^H \mathbf{e}\right\}$ is a monotone increasing function of $\mu$. Therefore, the optimal solution is $\boldsymbol{\theta}_{\text {opt }}\left(\mu^{\prime}\right)=e^{j \arg \left(\mathbf{d}+\mu^{\prime} \mathbf{e}\right)}$.

The complexity of the proposed MM-based solution is $\mathcal{O}\left(L N^3 \log _2\left(\Omega_{\max }-\Omega_{\min }\right) / \varepsilon\right)$, where $L$ is the number of iterations, and $\Omega_{\max }$, $\Omega_{\min }$ and $\varepsilon$ are the upper-bound, lower-bound, and accuracy of bisection search, respectively. In contrast, the complexity of the SDR-based solution is $\mathcal{O}\left(N^{6.5}\right)$ \cite{luo2010semidefinite}. Thus, the complexity of the MM-based solution is much lower than that of the SDR-based solution, especially for large-size IRS with a large value of $N$.
\section{Numerical Results}
\begin{figure}[!t]
	\centering
	\includegraphics[width=0.32\textwidth]{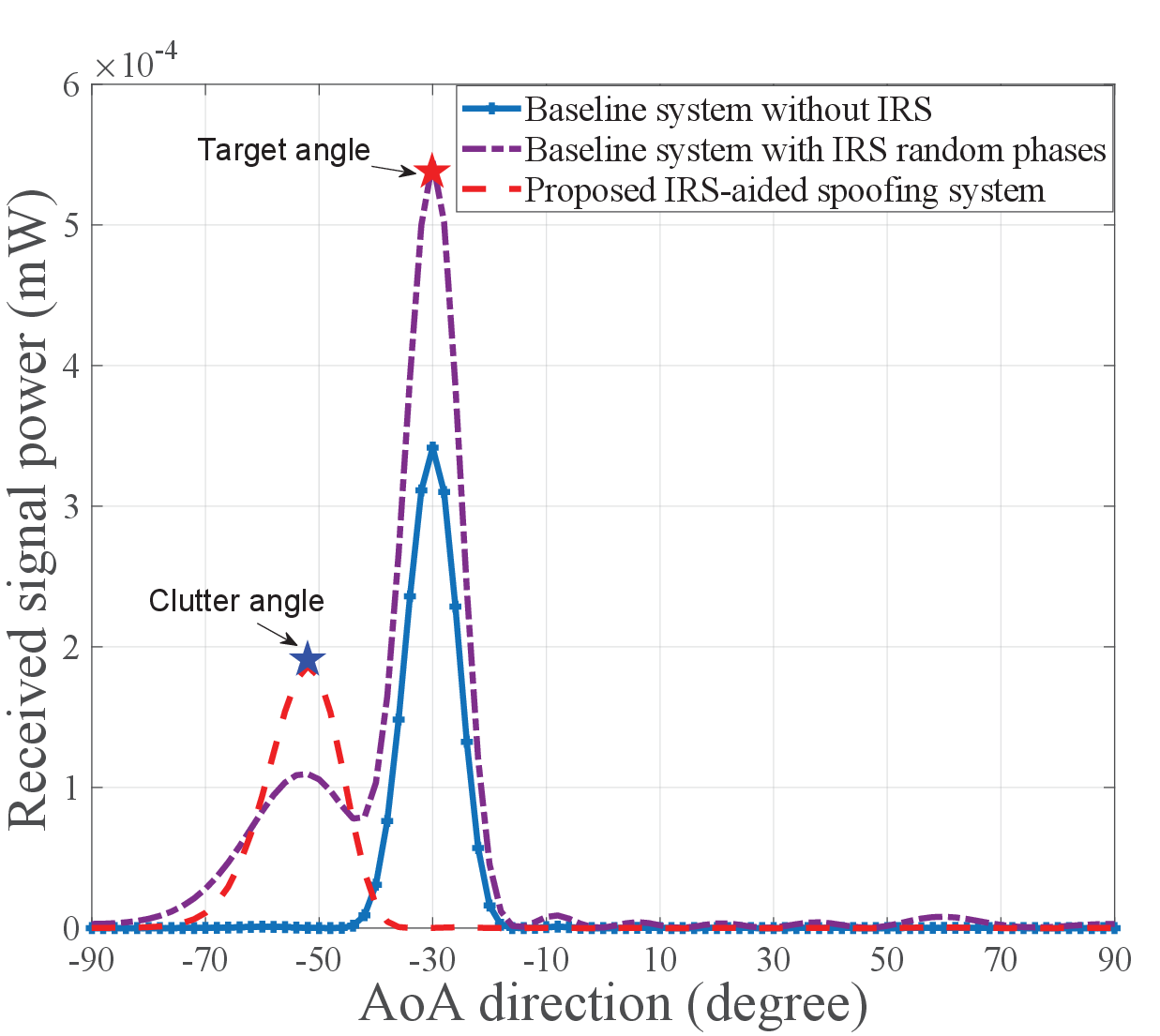}
	\caption{Adversarial radar's received power versus AoA direction}
	\label{fig:exp1}
\end{figure}
In this section, we evaluate the performance of the proposed IRS-aided radar spoofing system via numerical results. Two baseline systems are considered for comparison: 1) Without IRS, where the target is not equipped with IRS by setting $\boldsymbol{\theta}^{[t]}=\boldsymbol{0}$; and 2) IRS with random phase shifts, where $\theta_n^{[t]},n=1,\ldots,N$, follow the independent uniform distribution within $[0,2\pi)$. In our simulation, we consider the target, adversarial radar, and clutter are within the same 2D (vertical) plane by setting $\eta_{R \rightarrow I}^{[t]}=\eta_{R \rightarrow C}^{[t]}=\pi$ and $\eta_{C \rightarrow I}^{[t]}=0$. Thus, we only need to focus on the AoAs/AoDs $\left\{\vartheta_{R \rightarrow I}^{[t]}, \vartheta_{R \rightarrow C}^{[t]}, \vartheta_{C \rightarrow I}^{[t]}\right\}$. The mono-static radar is equipped with a UPA consisting of $M=8\times8=64$ transmit/receive antennas. The IRS is also assumed to be a UPA with $N=11\times11=121$ reflecting elements. The distances $d_{RI}^{[t]}$, $d_{RC}^{[t]}$ and $d_{CI}^{[t]}$ are set as 100 m, 97 m, and 36 m, respectively, which remain approximately (by ignoring the local distance perturbations) constant during the radar detection. Unless otherwise specified, the reference path gain at the distance of 1 m is set as $\alpha=-30$ dB for all links. The wavelength of the signal is set as $\lambda=0.05$ m. The RCS of the clutter is set as 7 dBsm.

In Fig. \ref{fig:exp1}, we plot the adversarial radar's received signal power over different AoA directions. It is observed that the target angle of $-30\degree$ is detected by the adversarial radar while the received signal power from the clutter angle of $-52\degree$ is negligible in the baseline system without IRS. The genuine target angle and the deceptive clutter angle can be both detected by the adversarial radar in the baseline system with IRS random phase shifts. In contrast, for the proposed IRS-aided radar spoofing system, the received signal power from the clutter direction is significantly enhanced for spoofing, whereas that from the true target direction is suppressed below a given detection threshold, $\gamma=10^{-7}$ mW. As a result, the adversarial radar will detect the clutter angle falsely as the target angle, and thus is spoofed successfully. 
\begin{figure}[!t]
	\centering
	\includegraphics[width=0.32\textwidth]{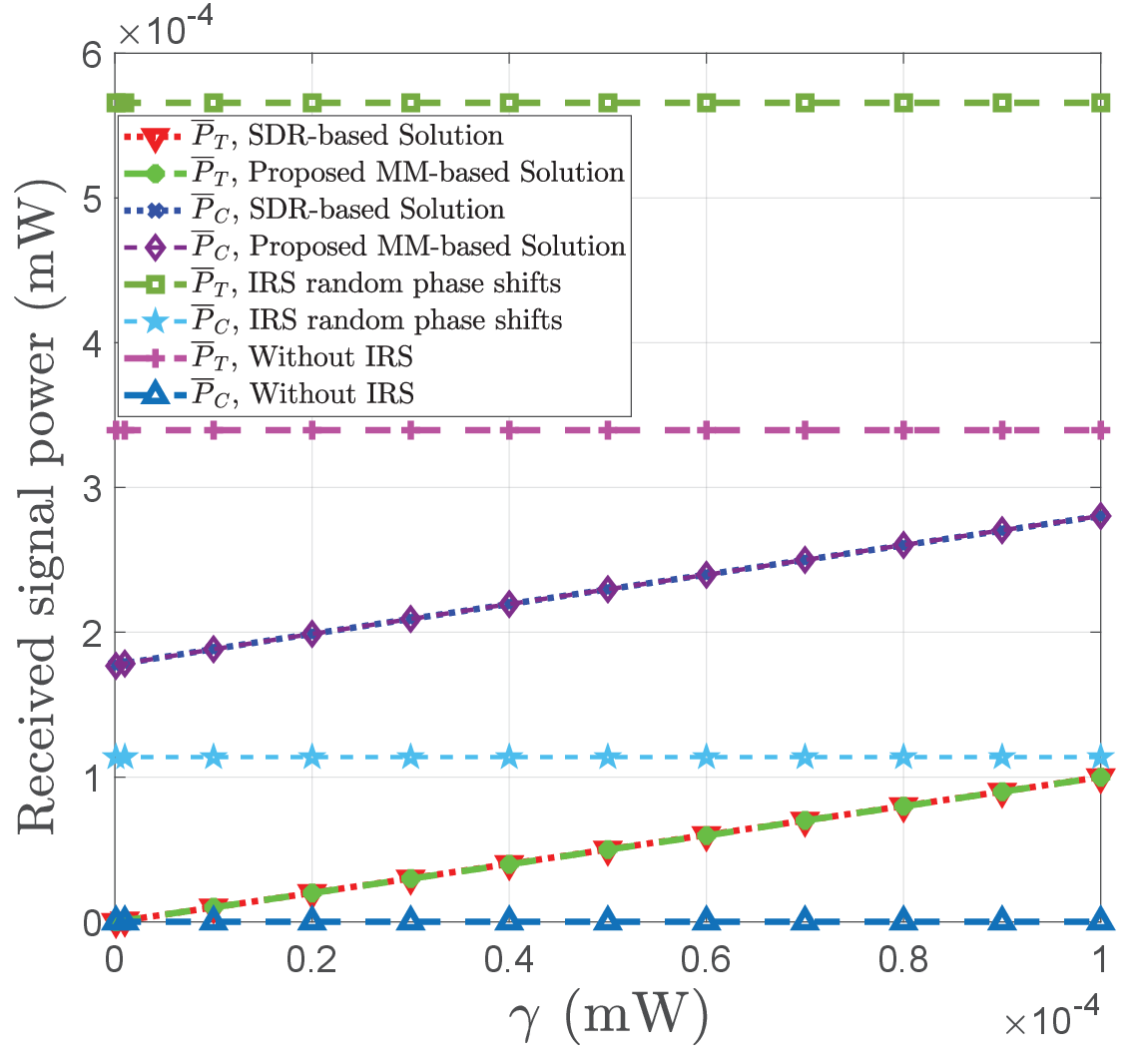}
	\caption{Received signal power at the adversarial radar from the target/clutter direction versus radar detection threshold, $\gamma$.}
	\label{fig:exp2}
\end{figure}
\begin{figure}[!t]
	\centering
	\includegraphics[width=0.32\textwidth]{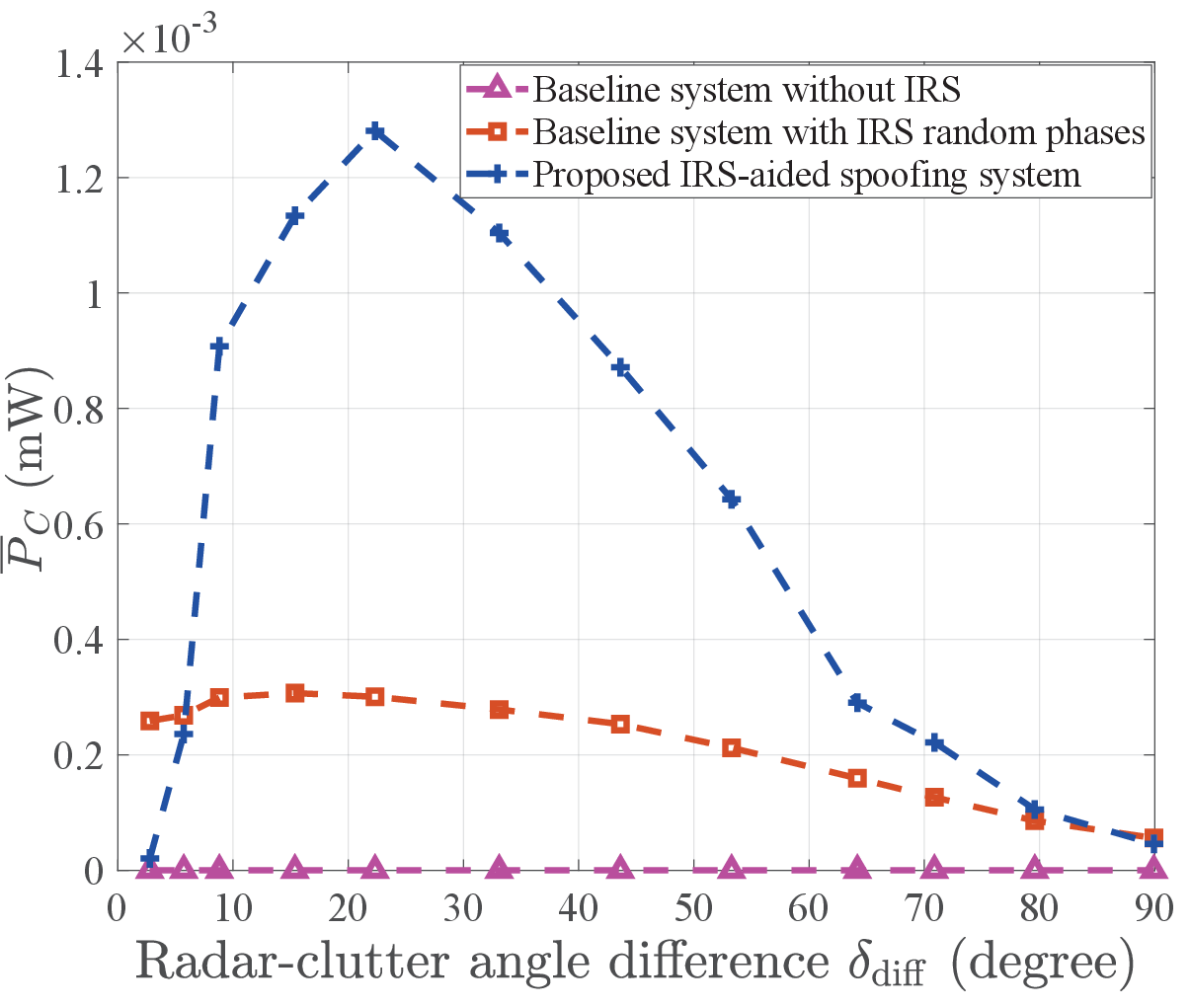}
	\caption{$\overline{P}_C$ versus radar-clutter angle difference, $\delta_\mathrm{diff}$.}
	\label{fig:exp3}
\end{figure}

In Fig. \ref{fig:exp2}, we examine the impact of radar detection threshold $\gamma$ on the received signal powers at the adversarial radar from the target and clutter directions, respectively. One can observe that increasing $\gamma$ is beneficial to spoofing (i.e., leading to higher $\overline{P}_C$) while it also makes the target less stealthy (i.e., leading to higher $\overline{P}_T$), thus resulting in a fundamental trade-off. In addition, the proposed MM-based solution achieves almost the same received power as the SDR-based algorithm, thus being appealing for practical implementation due to its much lower computational complexity.

In Fig. \ref{fig:exp3}, we investigate the impact of radar-clutter angle difference $\delta_{\mathrm{diff}}$ at IRS (i.e., $\delta_{\mathrm{diff}} = \vartheta_{R \rightarrow I} + \vartheta_{C \rightarrow I}$ when $\eta_{R \rightarrow I}= \pi$ and $\eta_{C \rightarrow I}= 0$, and $\delta_{\mathrm{diff}} =\vartheta_{C \rightarrow I} - \vartheta_{R \rightarrow I}$ when $\eta_{R \rightarrow I}= \eta_{C \rightarrow I}= \pi$) on the received power from the clutter direction, $\overline{P}_C$. Note that we move the clutter position only along the $x-$axis to change the value of $\delta_{\mathrm{diff}}$. For the proposed IRS-aided radar spoofing system, two interesting observations can be made as follows. First, $\overline{P}_C$ is relatively low when $\delta_{\mathrm{diff}}$ is smaller than $10\degree$, even when the product path gain (i.e. $\propto \frac{1}{d_{RC}^{[t]}d_{CI}^{[t]}}$) is relatively high. This is because it is difficult to design the IRS phase shifts to enhance $\overline{P}_C$ while suppressing $\overline{P}_T$ simultaneously due to the small radar-clutter angle difference. Second, $\overline{P}_C$ is also relatively low when $\delta_{\mathrm{diff}}$ is larger than $70\degree$, which is mainly due to the lower product path gain with the increase of $d_{RC}^{[t]}$ and $d_{CI}^{[t]}$. Therefore, it is practically wise to select a clutter with sufficient angle difference  $\delta_{\mathrm{diff}}$ as well as moderate distance from the target for effective radar spoofing.  
\section{Conclusion}
In this letter, we proposed a new IRS-aided radar spoofing strategy for deceiving the adversarial radar with a false clutter angle as well as concealing the true target angle at the same time. The IRS reflection phase shifts were designed to maximize the received echo signal power at radar from the clutter direction, subject to the constraint that the echo signal power from the target direction was lower than a given detection threshold. A low-complexity solution was proposed to solve this problem efficiently. Simulation results validated the effectiveness of the proposed radar spoofing system design as compared to baseline schemes.
\bibliographystyle{ieeetr}
\bibliography{reference}
\end{document}